\documentclass{aastex}
\usepackage{rotating}
\usepackage{natbib}
\usepackage{url}
\usepackage[usenames]{color}
\usepackage{hyperref}
\usepackage{epstopdf}
\usepackage{float}

\begin{document}
\nocite{*}

\date{September 2015}

\newcommand{\R}{\mathbb{R}}
\newcommand{\HST}{{\it HST }}
\newcommand{\JWST}{{\it JWST }}
\newcommand{\Spitzer}{{\it Spitzer }}
\newcommand{\Ktwo}{{\it K2 }}
\newcommand{\Kepler}{{\it Kepler }}
\newcommand{\TESS}{{\it TESS }}
\newcommand{\TBD}{{\bf TBD}}

\title{\Spitzer Observations of Exoplanets Discovered with The \Kepler \Ktwo Mission}

\author{Charles Beichman}
\affil{\it NASA Exoplanet Science Institute, California Institute of Technology, Jet Propulsion Laboratory}
\author{ John Livingston, Michael Werner, Varoujan Gorjian }
\affil{\it Jet Propulsion Laboratory, California Institute of Technology}
\author{Jessica Krick}
\affil{\it Infrared Processing and Analysis Center, California Institute of Technology}
\author{Katherine Deck, Heather Knutson, Ian Wong, Erik A. Petigura}
\affil{\it California Institute of Technology}
\author{ Jessie	Christiansen, David Ciardi}
\affil{\it NASA Exoplanet Science Institute, California Institute of Technology}
\author{Thomas P. Greene, Joshua E. Schlieder}
\affil{\it NASA Ames Research Center}
\author{Mike Line}
\affil{\it University of California, Santa Cruz}
\author{ Ian Crossfield}
\affil{\it  University of Arizona,  NASA Sagan Fellow}
\author{Andrew Howard, Evan Sinukoff}
\affil{\it Institute for Astronomy, University of Hawaii}

\section{Abstract}

We have used the {\it Spitzer Space Telescope} to observe two transiting planetary systems orbiting low mass stars discovered in the \Kepler \Ktwo mission. The system K2-3 (EPIC 201367065) hosts three planets while EPIC 202083828 (K2-26) hosts a single planet. Observations of all four objects in these two systems confirm and refine the orbital and physical parameters of the planets. The refined orbital information and more precise planet radii possible with \Spitzer will be critical for future observations of these and other \Ktwo targets.  For K2-3b we find marginally significant evidence for a Transit Timing Variation between the \Ktwo  and  \Spitzer\ epochs.

\section{Introduction}
\subsection{Demographics and Properties of Planets Orbiting  M Stars}

One of the primary goals of the re-purposed \Kepler spacecraft (the ``\Ktwo mission'') is a wider survey of late type stars than was achieved within the primary \Kepler mission \citep{Beichman2013, howell2014}. Population studies of the \Kepler data suggest an increased incidence of lower mass planets orbiting M stars \citep{Howard2012}. Although \Kepler observed only about 3000 M stars, initial results suggest a high incidence of planets orbiting low mass stars, approaching 100\% \citep{Dressing2013,Dressing2015}, of which up to 25\% may reside in the loosely defined stellar Habitable Zone (HZ) \citep{Dressing2015}. These effects must be explained in the context of planet formation theory \citep{Payne2007, Mordasini2012} and make the validation of these trends with a larger sample of great interest. By surveying a dozen or more fields, each containing $\sim$4,000 late type stars, \Ktwo promises to increase the sample of M star planetary systems more than tenfold.

 Planets orbiting M stars are important for reasons beyond their demographics. Because cool stars have smaller radii than earlier spectral types, the transit signal of a given sized planet is proportionately larger, resulting in easier follow-up spectroscopic observations with the {\it Hubble Space Telescope}, \HST\ \citep{Knutson2014}, and soon, the {\it James Webb Space Telescope}, \JWST\ \citep{Beichman2014}. While an Earth-analog (1 $R_\oplus$) orbiting a solar type star in a 1 AU HZ produces a 84 parts per million (ppm) transit signal every 365 days, the same planet orbiting in the HZ of an M3 star produces a $>$500 ppm signal every $\sim$30 days. Thus, \JWST\ spectroscopy will be able to probe down to at least the Super-Earth level \citep{Batalha2014} for late type stars. Stellar brightness is another critical parameter for transit spectroscopy. In this regard, \Ktwo offers an advantage over \Kepler by covering $\sim$10 times more sky so that with careful selection it will be possible to target M stars that are 1-2 mag brighter than those in \Kepler's primary field. Eventually, the \TESS mission with its all-sky coverage will gain an average 3-5 mag in host star brightness over \Kepler \citep{Ricker2014, Sullivan2015}. 
 
Numerous groups have proposed M star candidates for \Ktwo and are engaged in follow-up activities to identify, validate and characterize candidates found in the \Ktwo light curves. In this paper we introduce  a follow-up effort using the {\it Spitzer Space Telescope} to improve the orbital ephemerides and other properties of \Ktwo planets hosted by cool stars. We report the  results for two systems: K2-3, an M0 star with three planets (EPIC 201367065b,c,d; \citet{Crossfield2015}) and EPIC 202083828 (hereafter K2-26), an M1 star with a single transiting planet \citep{Schlieder2015}.

\subsection{The \Spitzer  \Ktwo Transit Program}

A proposal to follow-up planets hosted by M stars by \Ktwo using \Spitzer observations at 4.5 $\mu$m (IRAC Channel 2) was approved in Cycle 11 (Werner, PI; Program 11026). \Spitzer observations will augment and complement \Ktwo results in a number of important ways. 

\begin{enumerate}
\item \Kepler's 30 minute observing cadence means that the ingress and egress of a transit or even the entire transit, which might be as short as 1 hr for a late M star, will have only a handful of \Kepler samples per event. For comparable signal-to-noise ratio (SNR) on the transit depth, \Spitzer transits provide much tighter constraints on the orbital and system parameters because of the much finer sampling (0.4-to-30 sec vs. 30 minutes for \Ktwo). 
\item By observing a year or more after \Ktwo's measurements, \Spitzer in conjunction with the original \Ktwo results can dramatically improve the orbital ephemerides and thus enable accurate predictions of transit timing many years into the future, which will be particularly important for \JWST\ observations. In the case of multiple systems, \Spitzer may reveal Transit Timing Variations (TTVs) which may be used to estimate planetary masses.
\item M stars show strong limb darkening in the \Kepler bandpass which complicates the
determination of the transit parameters and the planetary characteristics. Much less limb darkening is present in the \Spitzer 3.6 and 4.5 um bands \citep{Claret2011}.
Thus the \Spitzer measurements permit a much cleaner determination of the transit parameters, particularly given the sampling issue discussed above.
\item To first order, the depth of a transit should be achromatic. This lack of change of transit depth with wavelength means that \Spitzer results can be used to reject certain false positive alternatives to the transit interpretation, e.g. a low mass stellar companion, particularly in the absence of radial velocity observations. There are, however, small wavelength dependent variations between the visible and infrared which may be interpreted in terms of atmospheric structure, e.g. the presence of molecular absorptions or a temperature inversion. Such claims are at the limits of \Spitzer's accuracy  \citep{Evans2015, Wong2015}. 
\item Two additional advantages of \Spitzer, a lower level of photospheric noise (``stellar jitter'') in the infrared  compared with visible wavelengths, and the detection of secondary eclipses for hot, short period planets, may eventually be demonstrated on M stars still to be identified by \Ktwo.
\end{enumerate}

Taking the above considerations into account, \Spitzer provides an important means of screening M star exoplanets and identifying those most promising for \JWST\ follow-up, especially because IRAC spans the middle of \JWST's spectroscopic wavelength range. The results presented herein will demonstrate the value of \Spitzer in all of these areas. The approved \Spitzer program will observe of order 30 transiting systems with over 450 hours of telescope time with the goal of improving planetary and orbital properties for the brightest, most promising targets for future spectroscopic follow-up.

\subsection{The Need for Improved Ephemerides}

A focus of this paper will be the importance of \Spitzer observations to improve significantly the ability to recover future transits. At the simplest level, the ability to predict the time of a future transit, $T(n)$, depends on the uncertainty in the reference time for an initial mid-transit time, $\sigma(T_0) $ and the uncertainty in the orbital period, $\sigma P$, projected $n$ orbits into the future:

\begin{equation}
T(n)=T_0+n P
\end{equation}
\noindent and 
\begin{equation}
\sigma(T_n)=\sqrt{\sigma(T_0)^2+(n\sigma(P))^2}
\end{equation}

Equation (2) shows that the uncertainty in the mid-transit time increases linearly with orbit number after the reference orbit. Thus, for example, for the planet K2-3d (see Table~\ref{K2data}) \Ktwo data alone yield $T_0$ (BJD) =2456826.2233 $\pm$0.0039 (2014 June 17) and $P=44.5629\pm0.0057$. By the time \JWST\ gets around to observing this object on, say, BJD 2458817.609 (2019 Dec 14), 45 orbits after its initial observation by \Ktwo, the 1 $\sigma$ uncertainty in the transit mid-point would be $\sim$6 hours. Such a large uncertainty would increase the duration required for a \JWST\ observation by $\sim$12 hours (from -1$\sigma$ to +1$\sigma$) relative to a short 4 hr transit to be sure of capturing the entire transit at even the 1 $\sigma$ level. As we demonstrate in this paper, the addition of even a single \Spitzer observation can reduce this uncertainty by a factor of 5-10.

\section{Observed Targets}

The first two objects we observed with \Spitzer came from early discoveries from  \Ktwo in  Fields 0 and 1\footnote{http://keplerscience.arc.nasa.gov/k2-fields.html}. 

The multiple system K2-3 (EPIC 201367065) has three planets orbiting an M0 star \citep{Crossfield2015} as described in Table 1. The three planets (b,c,d) have radii in the range of 1.2-2.4 R$_\oplus$ and were observed by \Ktwo with 8,4 and 2 transits, respectively. The star is bright in near-IR wavelengths ($K_s=8.56$ mag, \citet{Skrutskie2006}; WISE [4.6]= 8.42 mag, \citet{ Wright2010}) with a radius of 0.56$\pm$0.068 R$_\oplus$ making it a promising target for \JWST\ follow-up which requires bright targets for high SNR spectroscopy. The outermost planet, K2-3d, is in the nominal Habitable Zone with an insolation of 1.5 $\pm$ 0.5 times our Earth's and an effective temperature around 300 K \citep{Crossfield2015}.

Examination of the sources in the Campaign 0 field led to the identification of a 2.7 R$_\oplus$ planet with an effective temperature of $<$ 500K orbiting the M1 star K2-26 (K$_s=10.53$ mag; WISE [4.6]= 10.35 mag) on a 14.5 day period. As described in \citet{Schlieder2015} the host star is an M1.0$\pm$0.5 dwarf with near-solar metallicity, [Fe/H]=$-0.13\pm0.15$ and a stellar radius of 0.52 $\pm$ 0.08 R$_\odot$. \citet{Schlieder2015} argue that from the examination of HIRES spectroscopy, which rules out spectroscopic binaries earlier than M4.5V, LBT/LMIRcam and Robo-AO adaptive optics imaging, and archival survey images spanning more than 50 years, that the likelihood of this planet candidate being a false positive due to an eclipsing binary or hierarchical multiple system is extremely small. Thus we included this object in our \Spitzer program.

\section{Spitzer Data and Analysis}
In March 2015 K2-3b was observed by \Spitzer on two epochs while planets `c' and `d' were observed once each. All three were observed one more time each approximately 6 months later in Sept. 2015 (Table~\ref{SpitzerLog}). K2-26b was observed once in March 2015. The science observations were timed to begin 2 hours before the start of the transit and end two hours after the end of the transit to allow adequate baseline on either side of the event. We preceded the main observation with a 30-min pre-observation of the target to mitigate the effect of large drifts across the pixel due to temperature changes in the spacecraft after large slews from the preceding observations  \citep{Grillmair2012}.

All observations were obtained with \Spitzer IRAC Channel 2 (4.5 $\mu$m; \citet{Fazio2004, Werner2004}) using staring mode observations. We choose Ch2 because the dominant instrumental systematic of changing gain as a function of position is a smaller effect in Ch2 than Ch1 \citep{Ingalls2012}; the diminished effect of limb-darkening in Ch2 is also advantageous. Staring mode is standard practice for exoplanet observations in order to keep the star on one position within a single pixel. To ensure that the most well-calibrated position with minimal gain variation is achieved on the pixel, a peak-up star was used to place the target star (corrected for proper motion, Table~\ref{K2data}) on the ``sweet spot'' of the central pixel. Exposure times were chosen to maintain the well depth in the linear regime of the detector. K2-3 was observed in sub-array mode with 2s frame times and  K2-26b was observed with 12s exposures in full array mode. The same ``sweet spot" was used for both stars. This observing strategy resulted in over 53,928 individual photometry points used in the analysis described below.

\subsection{Photometric Analysis}


  Centroiding and aperture photometry were performed using the Python package {\it photutils}. Aperture photometry was computed for each exposure using radii ranging from 2.0 to 2.9 pixels in 0.1 pixel increments, as well as 3.0 to 5.0 pixels in 0.5 pixel increments. Optimal photometric radii were determined for each dataset by choosing the time series with minimal scatter, thus minimizing the contribution from background noise while including enough of stellar flux to maximize SNR. The typical radii values used were 2.2 or 2.3 pixels, which is consistent with independent analysis of optimal \Spitzer transit photometry \citep{Krick2015}. Sky background levels and photometric uncertainties were computed taking into account known characteristics of the detector. Because of the small size ($32\times32$ pixels) of sub-array images, this estimate entails a trade-off between good number statistics and contamination from the stellar PSF. We approached this  by fitting a Gaussian to each frame after masking pixels within the central PSF and central 2 rows and columns, as well as the top row, which is systematically biased to lower values (see \citet{Knutson2012} for more detailed discussion).

\subsection{Spitzer Systematics}

The largest systematic in \Spitzer photometry arises from intrapixel gain variations.  Spacecraft-induced motion coupled with an under sampled PSF lead to measured flux variations of order a few percent   \citep{Ingalls2012}. The spacecraft motions are of several types: variable-duration thermal settling of the spacecraft, pointing control errors resulting in long-term drift, a 39 minute sawtooth pointing oscillation due to the cycling of a battery heater in the spacecraft bus, as well as both high and low frequency jitter due to a variety of possible causes, including harmonic coupling of the reaction wheel assembly to the spacecraft structure \citep{Grillmair2012}.

We  used  the pixel-level decorrelation (PLD) method technique \citep{Deming15} to reduce these systematic photometric variations.  Similar to \citet{Deming15}, we fit the PLD pixel coefficients simultaneously with a temporal systematic model. We opt for a linear (instead of quadratic) ramp in time because the data do not obviously warrant the increase in model complexity, and a quadratic ramp is more likely to be degenerate with the transit signal. Thus the total deviation in signal at time $t$ is modeled as:

\begin{equation}
\Delta S^t = \sum\limits_{i=1}^N c_i \hat{P}_i^t + T(t) + m t + b
\end{equation}

where the $c_i$ are the coefficients that represent the partial derivatives from the Taylor expansion described by \cite{Deming15}, $T(t)$ is the transit signal, and $m$ and $b$ are the coefficients of the linear ramp in time. $\hat{P}_i^t$ is the $i^{th}$ pixel value of the normalized pixel grid at time $t$: 

\begin{equation}
\hat{P}_i^t = \frac{P_i^t}{\sum\limits_{i=1}^N P_i^t} 
\end{equation}

  We tried using both a $3\times3$ and a $5\times5$ pixel grid centered on the target, and we found that while a $3\times3$ grid produced good results, a $5\times5$ grid further reduced the residual RMS at minimal computational cost. We took an iterative approach to fitting the systematic coefficients which allows for a gradual refinement due to intermediate improvements in the estimate of the non-systematic signal, i.e. the transit. We fit the systematic coefficients while the transit parameters are held fixed, then fit for the transit parameters while the systematic coefficients are held fixed, and repeat until a convergence criterion is met. At each iteration the parameters are updated with their new maximum likelihood estimate (MLE) values using the Nelder-Mead simplex algorithm and a Gaussian likelihood.  

  Because the transit depths of these planets are in some cases (K2-3d) approaching the limit of \Spitzer's precision, simultaneous fitting of the systematic and astrophysical parameters can lead to the non-convergence of a wide variety of numerical optimization algorithms. Initial testing showed that a modest improvement in the estimate of the systematic component prior to fitting any transit parameters prevented this, which led to the development of the iterative approach described above. This is perhaps  due to operating near the limit of the instrument capability, combined with degeneracies between transit and systematic parameters which complicate the objective function. For example, tests conducted with higher SNR \Spitzer transit data in which the amplitude of the systematic signals is smaller compared to the transit signaltransit depth do not exhibit the same difficulty. However, this effect is mitigated by operating on un--binned data, and the transit parameter estimates derived from the \Ktwo data are of sufficient quality.

\subsection{Transit Fitting and Derived Parameters}

In the context of the above discussion, one possible drawback of this iterative method in lower SNR datasets is an increased reliance on good starting guesses for the transit parameters, as these are held fixed while the initial PLD coefficients are fit. Although variation of bin size typically resulted in a tight range of fitted transit parameters, in some cases larger bin sizes resulted in increased sensitivity to initial parameter estimates. Testing with un-binned data showed that the sensitivity to initial parameter estimates is typically about two orders of magnitude smaller than the uncertainties derived from the final posteriors. Thus, in order to be more robust to uncertainties in the initial transit parameters, and because of the relatively low computational complexity of PLD, we analyzed the data without binning. Furthermore, although we typically detect no significant correlated noise (at 95\% confidence) after PLD, even low levels of residual correlation could induce biases in fits to the binned data. The iterative approach described above typically adds only minor additional complexity, so the bulk of the total computation cost is expended during the sampling of transit parameter posterior distributions.

For transit parameter estimates and uncertainties, we use the open-source {\it emcee} package \citep{Foreman2013}, an efficient Python implementation of the affine-invariant Markov Chain Monte Carlo (MCMC) ensemble sampler \citep{Goodman2010}. To fit the PLD-corrected data produced by the iterative method described above, we use the open-source PyTransit code \citep{Parviainen2015} to generate the transit model. For the planets of K2-3 we take a conservative approach to fitting the astrophysical parameters by using wide flat priors on the mid-transit time, scaled semi-major axis, inclination, and planet to star radius ratio. This produced fairly Gaussian posteriors for all parameters except the scaled semi-major axis, which has a distinctly skewed posterior due to degeneracy with the inclination parameter. For  K2-26b we use Gaussian priors set by the values reported by \citep{Schlieder2015}, because the egress of the transit in the \Spitzer data was too close to being missed to ensure good fits using flat priors. We ensured that the MCMC chains produced by the sampler were of sufficient quality by monitoring both the auto-correlation time and the acceptance fraction, as well as by visual inspection of the chains and corner plots of the posteriors. 

\begin{figure}[t!]
\includegraphics[scale=0.5]{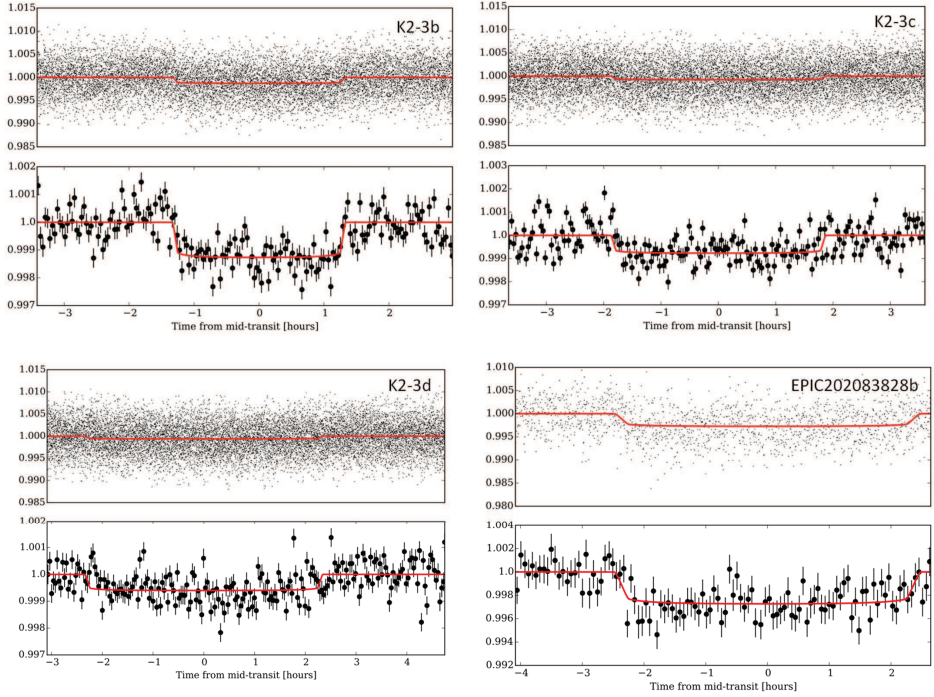}
\caption{\small\it Calibrated light curves for the 4 planets shown as normalized flux as a function of time from mid-transit.  The top plot of each of the 4 panels shows unbinned fluxes and the bottom plot shows binned fluxes, both PLD-corrected as described in the text.  Red lines are the best fit models.  Note the Y-axis change from un-binned to binned plots.Transit fits to \Spitzer data from March 2015 as described in the text. Top,left) Planet K2-3b; top,right) planet K2-3c; bottom,left) planet K2-3d; bottom, right) planet EPIC20203828b. \label{SpitzerPlots}} 
\end{figure}

The results of the correction of the \Spitzer photometry to reveal the transit signals are shown in Figure ~\ref{SpitzerPlots}. The derived system parameters are given in Table ~\ref{SpitzerFits}.

\section{Results and Discussion}

\subsection{Comparison of \Ktwo and \Spitzer Parameters}

The fundamental transit parameters of period and depth are consistent between the \Ktwo and \Spitzer datasets. Consider the case of K2-3b for which the \Spitzer data alone yield a period of 10.054440$\pm$0.0000053 days which can be compared with the \Ktwo-only value of 10.05402$\pm$0.00026 d \citep{Crossfield2015} which differ from one another by 1.6 $\sigma$.  By  combining the individual \Ktwo and \Spitzer transit times simultaneously, we obtain  a new, more accurate period (\citet{Crossfield2015} and Table~\ref{SpitzerFits}) of 10.0545435$\pm$0.000029 d (Table~\ref{Periods}) which differs from the \Ktwo-only value by 2$\sigma$. We discuss the apparent difference between the two estimates below in a discussion of possible Transit Timing Variations $\S$~\ref{sec:TTVs}.

\begin{figure}[t!]
\includegraphics[scale=0.75]{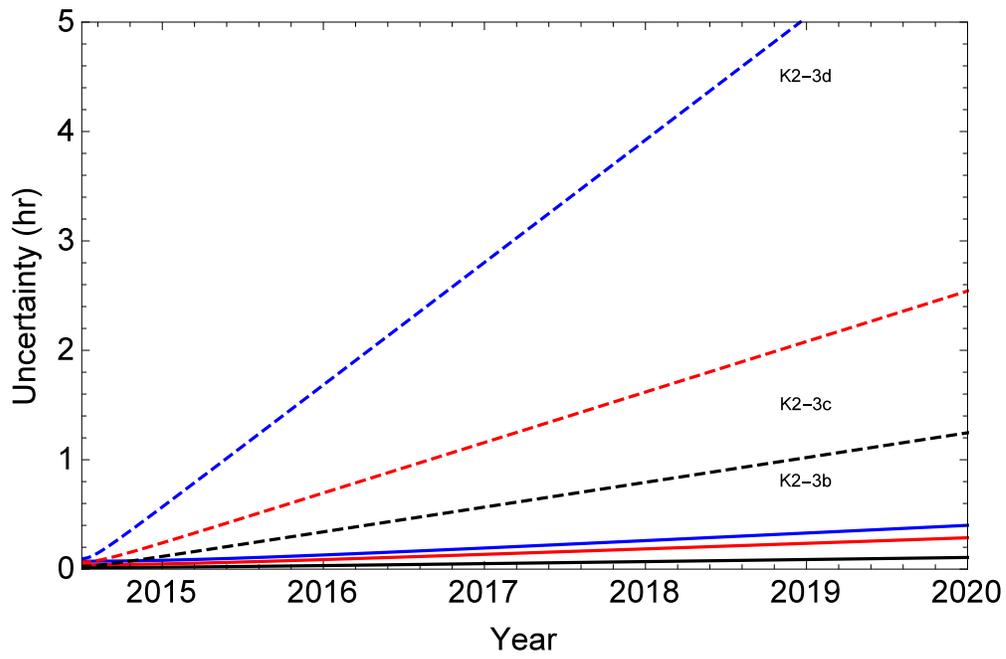}
\caption{\small\it Dashed lines represent uncertainties in projected transit times (ignoring possible Transit Timing Variations) based solely on \Ktwo data. Solid lines represent uncertainties from combined \Ktwo and \Spitzer data. The colors correspond to K2-3b (black), K2-3c (red) and K2-3d (blue). \label{uncs}} 
\end{figure}

In Table~\ref{Periods} and Figure~\ref{uncs} we present combined period estimates which represent a 5- to 10-fold improvement due to the longer temporal baseline in the combined datasets. Using these new values {\it and assuming no TTVs} we can project the ephemerides of K2-3b,c,d into the \JWST\ era ({\it circa} 2019 Dec) to be less uncertain than 0.11, 0.28, and 0.40 hr, respectively, compared with \Ktwo-only uncertainties of 1.3, 2.5 and 6 hr. These results show the importance of \Spitzer observations in greatly reducing the uncertainties in transit times for future observations.

The other parameter of primary importance is the depth of the transit, or the derived parameters R$_p$/R$_*$ and R$_p$/R$_\oplus$. These are very similar between the \Ktwo-only and \Ktwo-\Spitzer values for all four planets. As shown in Tables~\ref{K2data} and \ref{SpitzerFits}, the differences in the transit depths are within 1 $\sigma$ with refined values given in Table~\ref{Combined}. In the case of K2-26b it is important to note that the close similarity in transit depth provides further confidence in the planetary nature of the transiting source. False positives due to an eclipsing binary or hierarchical system would have a markedly different eclipse depth in the \Spitzer band \citep{Desert2015}.

\subsection{Transit timing}\label{sec:TTVs}

Deviations from the transit times predicted by a constant period (Keplerian) orbit can be indicative of mutual gravitational interactions between planets. TTVs have proven to be an important method of validating Kepler planets  and  measuring  dynamical masses (e.g. \citealt{Steffen2013} and \citealt{Carter2012}).

With the longer baseline allowed by \Spitzer, we examined the central transit times for the K2-3 system to assess the statistical significance of any transit timing variations.
In addition to measuring the transit times of the transits observed with \Spitzer (Table~\ref{centertimes}), we obtained  \Ktwo  lightcurves for all of the  individual K2-3 transit events \citep{Crossfield2015}  which allows us to fit individual transit times (Table~\ref{centertimes}) for the planets of K2-3 in the original \Ktwo  observing campaign. Note, however, that for any individual \Ktwo  transit there are only a handful of individual data points with which to fit each transit, with the result that the individual timing uncertainties are large, e.g. $\pm$2.5 minutes.

\begin{figure}[t!]
$
\begin{array}{c} 
 \includegraphics[scale=0.5]{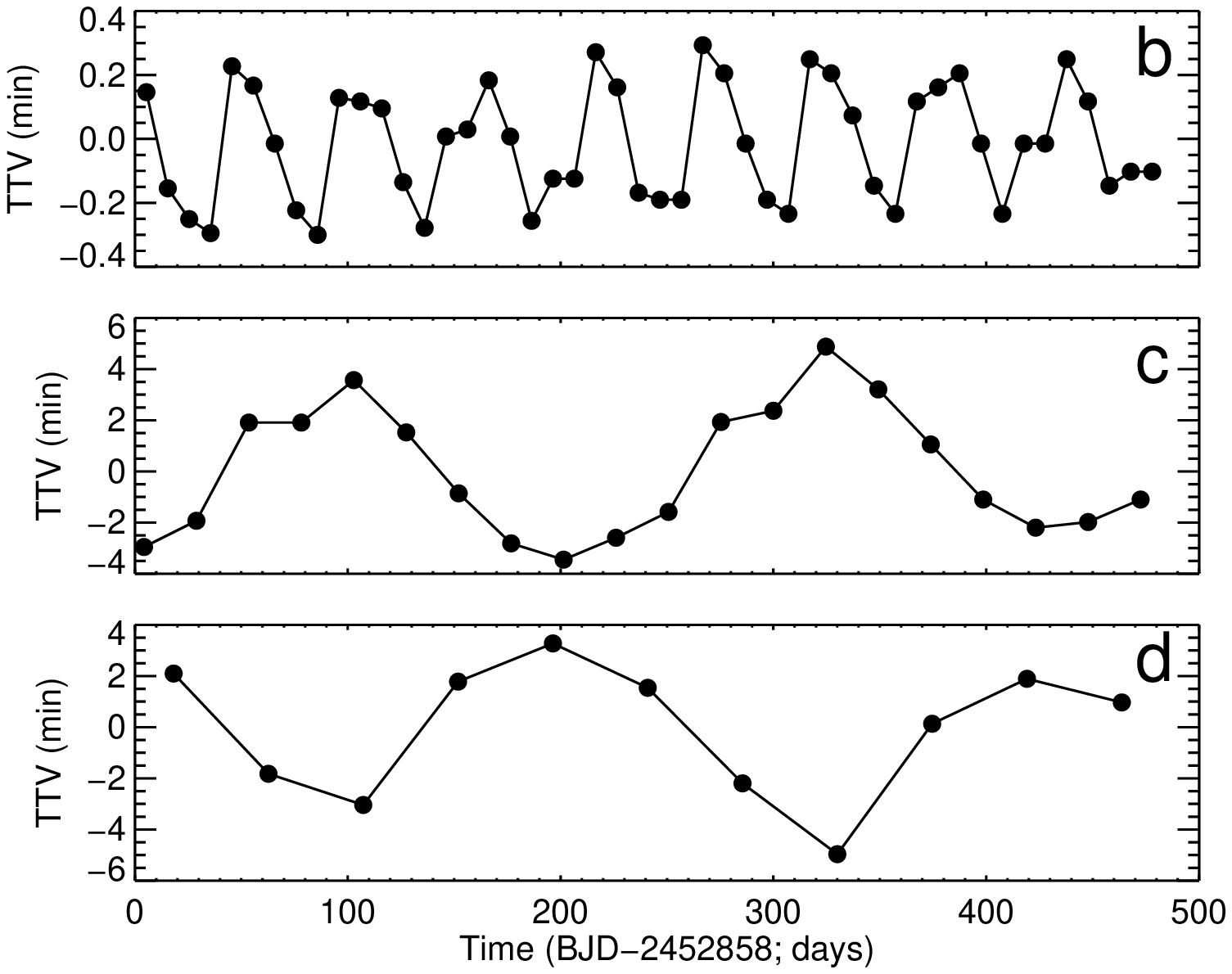}\\
 \includegraphics[scale=0.5]{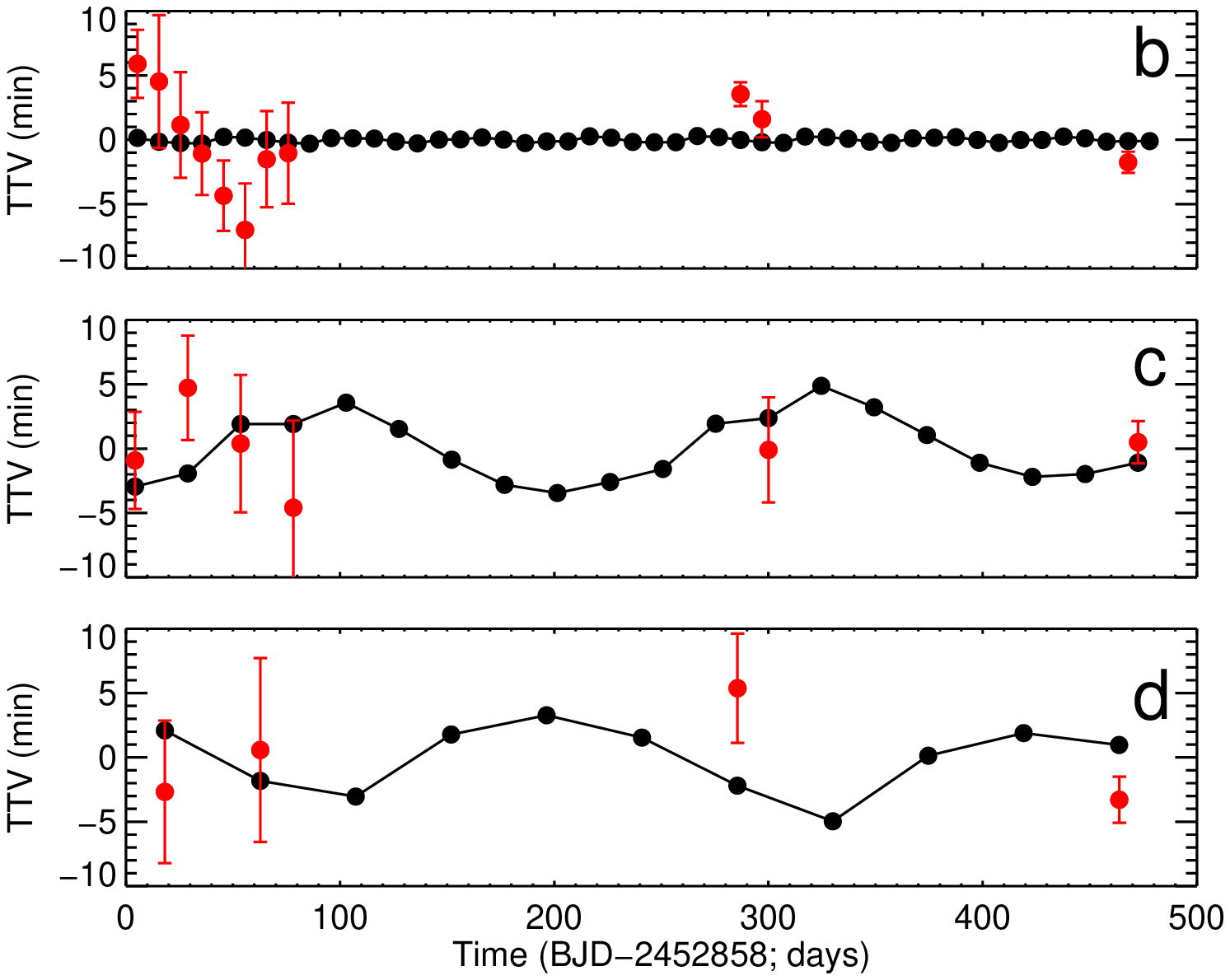}\\
 \end{array}
 $
\caption{\small\it top) Predicted TTVs for the three planets orbiting K2-3 based on an illustrative model assuming circular orbits and nominal masses. The predicted TTVs for K2-3b are only a fraction of a minute, much smaller than the observed deviations. Deviations for the outer planets could be as large as 5-10 minutes, but are not constrained by these data. bottom) The same models but with the data overplotted on an expanded scale. \label{TTV2}} 
\end{figure}

While the \Ktwo transit times for K2-3b alone are consistent with a constant period model (which is in turn consistent with the mean ephemeris reported by \citealt{Crossfield2015}), the combined \Spitzer and \Ktwo dataset show some evidence of TTVs.  When all 11 data points are considered the deviation from a constant period model is  statistically significant with a $\chi^2$ of 23 for 9 degrees of freedom.  Visual inspection of the individual \Ktwo  transit times of K2-3b suggest a coherent variation of $\sim$ $\pm$5 minutes relative to the mean period estimated from the combined dataset (shown in red, right upper panel of Figure~\ref{TTV2}). However, the major contributor to the $\chi^2$ is the \Spitzer data which suggest the existence of a  TTV with an amplitude around 2$\pm$1 minute. The individual transit times for the combined  \Ktwo+\Spitzer data for the outer two planets are consistent with constant period orbits (shown in red, lower two panels on right Figure~\ref{TTV2}) with $\chi^2=3.4$ and 1.4 with 4 and 2 degrees of freedom, respectively.

 

We consider briefly the amplitude of the TTVs one might expect for the K2-3 system in one representative configuration with very nearly circular orbits, a mass for K2-3b of $\sim 8M_\oplus$ \citep{Almenara2015}, and masses of $5M_\oplus$ for the outer two planets (based roughly on the mass-radius relationship in \citet{Weiss2014}). Illustrative TTVs for `b' and the outer two planets in this configuration calculated using the code {\it TTVFast} \citep{Deck2014} are shown in Figure ~\ref{TTV2}. The right panel of the figure shows how this model (in black) compares with the observed TTVs (in red). This model does not represent a fit to the data, it merely demonstrates a realistic possibility for the TTVs.

While the observed transit times of `c' and `d' are consistent in amplitude with those predicted by the simple model, the statistically significant $\sim 2$ minute deviations observed for K2-3b in the \Spitzer data are not. The very low amplitude of the model TTVs of `b' are due to the low masses of the perturbing planets, the wide separation between the `b' and the other planets, and the lack of proximity to low order mean motion resonance. 

To increase the amplitude to match the observations, our only truly unconstrained parameters are eccentricities and longitudes of pericenter, because the mass of `c' cannot be much larger than our estimated $5 M_\oplus$ (a mass of $\sim6.5 M_\oplus$ results if the planet density was equal to that of iron). Larger amplitude TTVs therefore require the orbits of `b' and `c' to be eccentric. However, since planet `c' interacts much more strongly with planet `d' than it does with `b', because the semi-major axis ratio is smaller between this pair compared with the (b,c) pair, this scenario would predict larger TTVs for `c' and `d' compared with those of `b'. 


In summary, we find tantalizing  evidence for TTVs for K2-3b between  \Ktwo  and \Spitzer epochs. The TTV amplitudes predicted for a configuration with nearly circular orbits and realistic masses is low enough to be consistent with a null detection of TTVs. More observations are required to assess whether  TTVs of a few minutes seen in the combined  \Ktwo  plus and \Spitzer data for planet `b', which are marginally  significant ($\sim 2 \sigma$) at this stage, are in fact reflective of stronger dynamical interactions. 

\citet{Schlieder2015} note the possibility that  K2-26b might have a non-zero eccentricity based on its transit duration ($\epsilon>0.14 \, 2\sigma$) The nonzero eccentricity, if real,  may indicate past or present interactions with a perturbing body. With the current data, we find that the observed transit times do not show TTVs at any significance. However, this does not necessarily rule out a nearby
perturbing planet, since the amplitude, timescale, and phase of a TTV signal depends on many unknown parameters.

\subsection{Prospects for \JWST\ observing}

The importance of objects like K2-3 with its system of 3 planets is driven by the desire to carry out spectroscopic observations with \JWST\ of planets in the size range of 1-2 R$_\oplus$. Such observations are possible for planets orbiting bright, late-type stars which yield a deep transit along with copious stellar photons to yield high(er) SNR spectroscopy. Along with GJ1214b, Kepler 138bcd and the newly discovered GJ1132b \citep{Berta2015}, K2-3 represents a  planetary system well suited for early \JWST\ spectroscopy follow-up, at least until TESS targets become available.  K2-3's planets have 1-2 $R_\oplus$, transit depths $>$100 ppm and a host stellar magnitude [4.6 $\mu$m]$\leq$ 8.4 mag. Indeed, K2-3 is almost a magnitude brighter than Kepler-138 (WISE [4.6 $\mu$m]$<$ 9.4 mag). Situated in the Habitable Zone, K2-3d will allow study of a temperate ($\sim$ 300 K) mini-Neptune or Super-Earth sized planet.

\begin{figure*}
$
\begin{array}{c} 
 \includegraphics[scale=0.5]{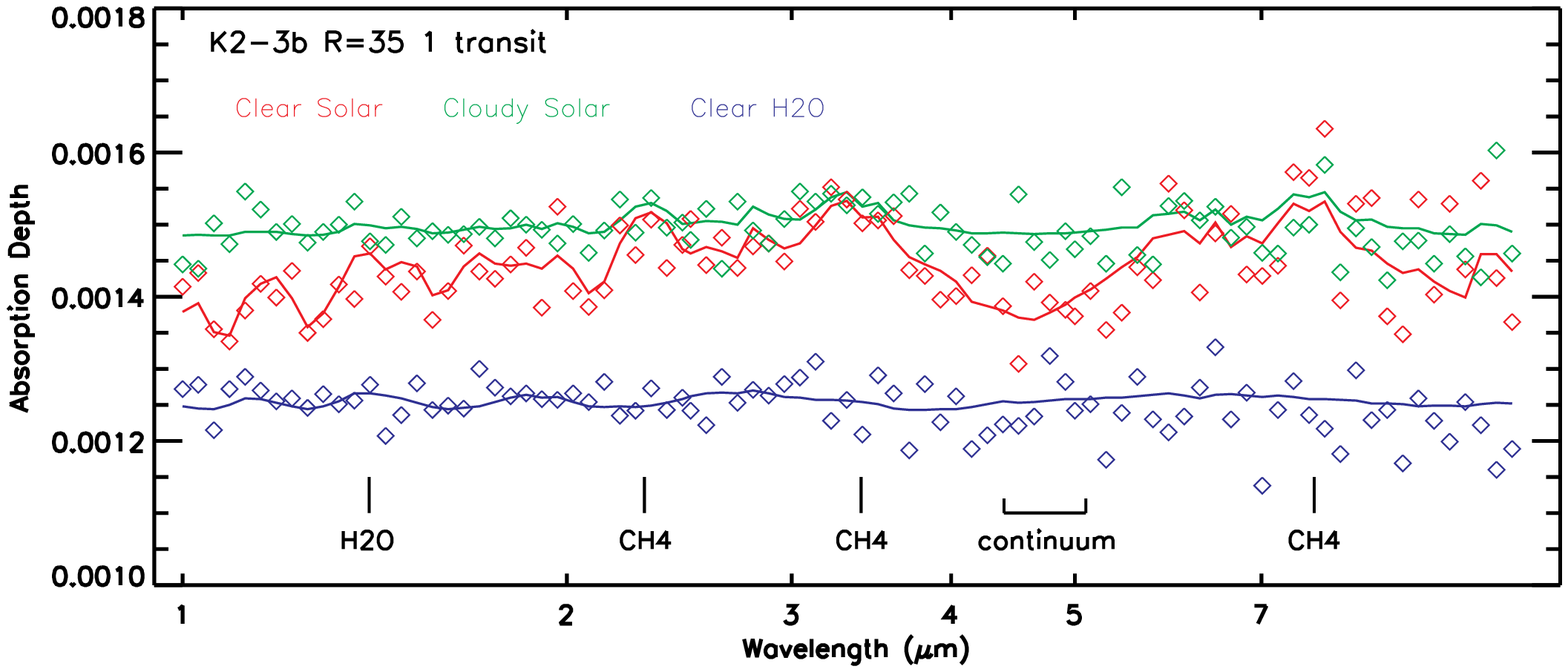}\\
 \includegraphics[scale=0.5]{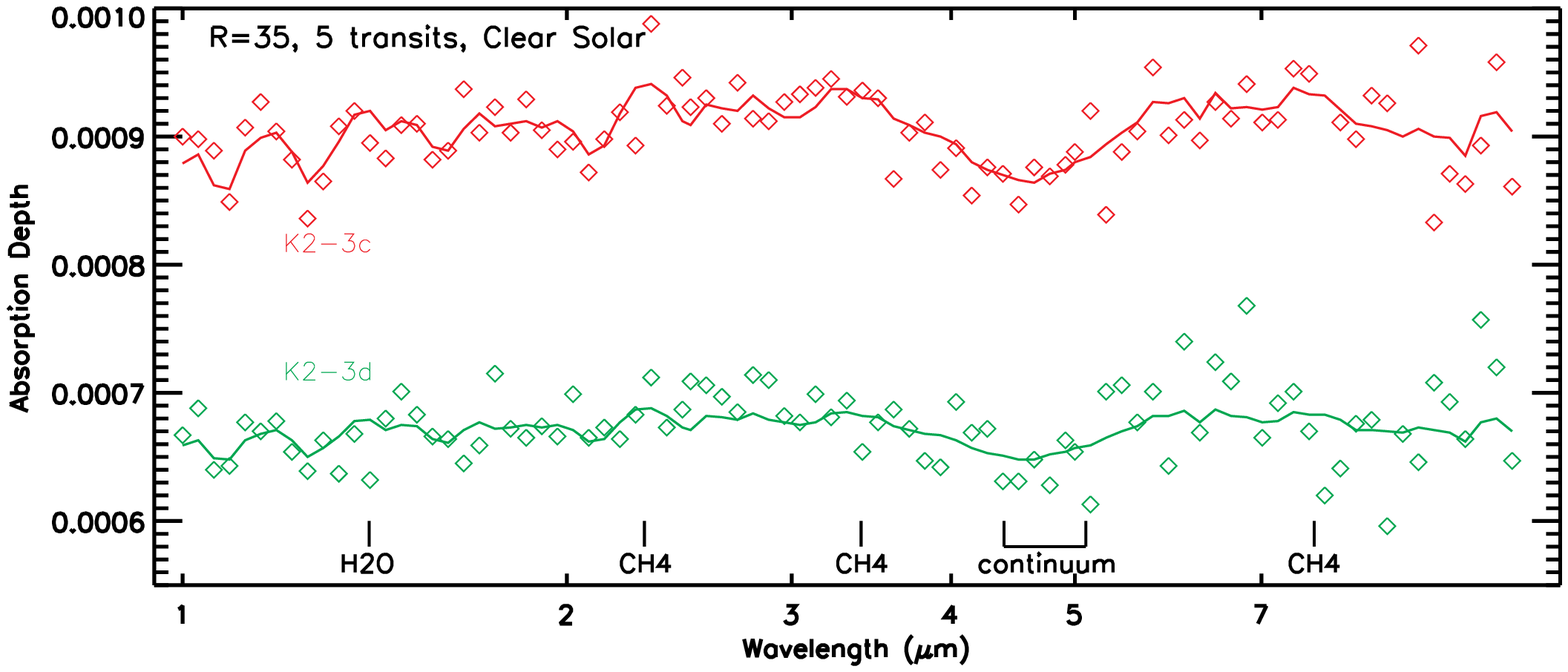}\\
 \end{array}
 $
 \caption{Top: Model and simulated \JWST NIRISS SOSS, NIRCam grism, and MIRI LRS transmission spectra of K2-3b, adapted from \citet{Greene2015}.Model atmospheres binned to R=35 are shown as continuous curves, and the
simulated data points (diamonds) were drawn from a Gaussian noise distribution for a single 2.3 hr transit plus an equal amount of time on the star at each wavelength. Photon noise and a systematic noise floor of 18 ppm ($1 \leq \lambda \leq 5$ $\mu$m; NIRISS and NIRCam) or 30 ppm ($\lambda > 5$ $\mu$m) were included in the total noise estimate. Bottom: Similar models for the planets K2-3c and d with data co-added for 5 transits at each wavelength. In both cases, a complete spectrum of the entire $1 - 11$ $\mu$m wavelength region will require observations of 3 or 4 transits with different instrument modes to obtain 1 transit at each wavelength. Strong features of H$_2$O, CH$_4$, and a clear continuum region are indicated. \label{JWST}}
 \end{figure*}

We developed thermochemical equilibrium models and simulated JWST observations of the transmission spectra of several possible atmospheres for K2-3b and clear solar composition atmospheres of K2-3c and K2-3d using the techniques employed by \citet{Greene2016}. Clear solar composition, cloudy solar, and 100\% H$_2$O atmospheres were generated for K2-3b, and we used the CHIMERA forward model \citep{Line2013a,Line2013b} and \citep{KBD14a, Kreidberg2014b} to generate the transit transmission spectra for all planetary atmospheres over $1-11 \mu$m. 
We simulated \JWST\ NIRISS SOSS ($1-2.5 \mu$m), NIRCam grism ($2.5-5.0 \mu$m), and MIRI LRS ($5.0-11 \mu$m) transmission spectra of the three planets in the K2-3 system using the techniques described in \citet{Greene2016}. The resultant spectra include photon, background, detector, and systematic noise components. They were binned to spectral resolving power R=35 and are shown in Figure~\ref{JWST}.   Noise levels were set by a combination of photon noise and a floor set by residual detector artifacts. We adopted  18 ppm noise floors for NIRISS and  NIRCam and 30 ppm for MIRI's longer wavelength detectors  \citep{Beichman2014}. The single transit measurement for  K2-3b   has a  total noise level between around 23-30 ppm at wavelengths less than 4 $\mu$m. The K2-3c and 3d simulations are for 5 transits, so that the noise floor dominates  with total noise 19 - 25 ppm from 1 - 5 $\mu$m before jumping up to 34 - 44 ppm at MIRI wavelengths.

Figure~\ref{JWST} shows that \JWST\ should detect strong H$_2$O (e.g., 1.4 $\mu$m) and the strong 2.3, 3.4, 7.7 $\mu$m CH$_4$ molecular features in the clear solar atmosphere models. Combining the information in a complete 1 - 11 $\mu$m spectrum (observed in 3 or 4 transits total) of K2-3b would allow distinguishing between clear solar and cloudy or high mean molecular weight (e.g., pure H$_2$O) atmospheres for that planet. More than 1 transit may need to be observed at each wavelength in order to measure the mixing ratios of detected molecules with moderate precision (better than 1 dex) if K2-3b does not have a totally clear solar composition atmosphere. Spectra from several transits will need to be co-added to detect these features individually in clear solar atmospheres for K2-3c and K2-3d.

\section{Conclusions}

We have used the \Spitzer {\it Space Telescope} to observe transits of the three planets orbiting the M0 star EPIC201367065 (K2-3b,c,d) and one planet orbiting the M star EPIC 202083828 (K2-26). The results allow us to refine the parameters of these planetary systems and greatly improve the precision of the ephemerides in support of future observations, notably with \JWST\ for spectroscopic follow-up. The observations have reduced the uncertainties in predicted transits of K2-3 planets from 4-6 hours down to less than 1 hour. For K2-26b, the observations provide strong support for the exoplanet nature of the transiting object by eliminating a number of false positives.

Predicted spectra for the K2-3 planets suggest that \Ktwo will provide  {\it JWST} with planets straddling the super-Earth to mini-Neptune classes suitable for spectroscopic characterization early in  {\it JWST's} mission. 

These observations represent just the initial results of a larger \Spitzer program which will improve the ephemerides, search for Transit Timing Variations, and in some favorable cases, provide observations of secondary eclipses.

\section{Acknowledgements}
This research has made use of data from the Infrared Processing and Analysis Center/California Institute of Technology, funded by the National Aeronautics and Space Administration and the National Science Foundation. This work was based on observations obtained with numerous facilities: the \Spitzer Space Telescope, which is operated by the Jet Propulsion Laboratory, California Institute of Technology under a contract with NASA; the Two Micron All Sky Survey, which is a joint project of the University of Massachusetts and the Infrared Processing and Analysis Center/California Institute of Technology, funded by the National Aeronautics and Space Administration and the National Science Foundation; and the Wide-field Infrared Survey Explorer, which is a joint project of the University of California, Los Angeles, and the Jet Propulsion Laboratory/California Institute of Technology, funded by the National Aeronautics and Space Administration. We also took advantage of the NASA Exoplanet Archive. Some of the research described in this publication was carried out in part at the Jet Propulsion Laboratory, California Institute of Technology, under a contract with the National Aeronautics and Space Administration. Ian Crossfield was funded by NASA through the Sagan Fellowship Program executed by the NASA Exoplanet Science Institute. Copyright 2014 California Inst of Technology. All rights reserved.

\clearpage

\begin{deluxetable}{lccc|cccc}
\tablecaption{Properties of \Ktwo Stars and Planets \label{K2data}}
\tabletypesize{\scriptsize}
\tablehead{Star &Spec &Kepler Mag & K$_s$ &Planet& R$_{pl}$&R$_{pl}/R_*$&Period \\
	 Name/Position &Type &	 (mag) &(mag) & &(R$_\oplus$)& &(day)}
\startdata
K2-3$^1$& M0$\pm$0.5& 11.57&8.56 &b &2.14$\pm$0.27&0.0348$^{+0.0012}_{-0.0007}$ &10.05402$\pm$0.00026\\
 EPIC201367065 & & &	 &c &1.72$\pm$0.23&0.027$^{+0.0014}_{-0.0008}$ &24.6454$\pm$0.0013\\
\multicolumn{3}{l}{11$^h$29$^m$20$^s$.49\, 	-01$^o$27$^\prime$18$^{\prime\prime}$.4 (J2000, Epoch 2015.7)$^3$} &	 &d &1.52$\pm$0.20&0.025$^{+0.0014}_{-0.0017}$ &44.5629$\pm$0.0057\\ \hline
K2-26$^2$&M1.0$\pm$0.5&12.47&10.53&b&2.67$^{+0.46}_{-0.42}$&0.0471$^{+0.0037}_{-0.0021}$ &14.5665$^{+0.0016}_{-0.0020}$\\
\multicolumn{3}{l}{06$^h$16$^m$49$^s$.55\, 	+24$^o$35$^\prime$45$^{\prime\prime}$.0 (J2000, Epoch 2015.7)$^3$} &	 & & &\\
\enddata
\tablecomments{$^1$Crossfield et al 2015. $^2$ \citet{Schlieder2015}.$^3$
The positions of both stars were corrected for proper motion from Epoch 2000 to the date of observation using positional data from the WISE and 2MASS surveys.}
\end{deluxetable}

\begin{deluxetable}{lccc}
\tablecaption{Spitzer Observing Log\label{SpitzerLog}}
\tablehead{Planet &Spitzer &Exposure &Observing \\
	 Name &Duration (hr) &Time (sec) & Date (UT) }
\startdata
K2-3b &7.0	& 2&2015 March 13 \\
K2-3b &7.0	& 2&2015 March 23 \\
K2-3c &7.9	& 2&2015 March 26	 \\
K2-3d &8.5	& 2&2015 March 11 \\
K2-3b &7.0	& 2&2015 September 10 \\
K2-3c &7.9	& 2&2015 September 15	 \\
K2-3d &8.5	& 2&2015 September 06 \\
K2-26b& 7.4& 12&2015 March 12\\
\enddata
\end{deluxetable}

\begin{deluxetable}{lc|c}
\tablecaption{Transit Parameters Derived from Fits to the \Spitzer Data\label{SpitzerFits}}
\tabletypesize{\scriptsize}
\tablehead{Parameter &Parameter&Parameter}
\startdata

Object&K2-3b (2015 March 13)&K2-3b (2015 March 23)\\
SMA/R$_*$&29.60$^{+1.44}_{-4.92}$&30.43$^{+2.68}_{-6.78}$\\
Incl (deg)&89.71$^{+1.03}_{-1.00}$&89.31$^{+1.25}_{-1.25}$\\
R$_p$/R$_*$&0.0346$^{+0.0007}_{-0.0007}$&0.0371$^{+0.0008}_{-0.0007}$\\
T0 (BJD)&2457094.94852$^{+0.00069}_{-0.00071}$&2457105.00231$^{+0.00061}_{-0.00071}$\\

Object&K2-3b (2015 Sept 15)&\\
SMA/R$_*$&30.79$^{+2.40}_{-7.10}$\\
Incl (deg)&90.19$^{+1.20}_{-1.41}$\\
R$_p$/R$_*$&0.0341$^{+0.0008}_{-0.0007}$&\\
T0 (BJD)&2457275.92813$^{+0.00091}_{-0.00069}$&\\

Period (Spitzer Only)&\multicolumn{2}{c}{10.0544403$\pm$0.0000530\,d}\\\hline

Object&K2-3c (2015 March 26)&K2-3c (2015 Sept 6)\\
SMA/R$_*$&51.12$^{+5.81}_{-11.98}$&50.71$^{+4.55}_{-10.40}$\\
Incl (deg)&89.98$^{+0.89}_{-0.92}$&89.86$^{+0.80}_{-0.78}$\\
R$_p$/R$_*$&0.0270$^{+0.0008}_{-0.0007}$&0.0271$^{+0.0009}_{-0.0010}$\\
T0 (BJD)&2457108.03149$^{+0.00321}_{-0.00239}$&2457280.55982$^{+0.00217}_{-0.00211}$\\

Period (Spitzer Only)&\multicolumn{2}{c}{24.6469050$\pm$0.0005002\,d}\\\hline

Object&K2-3d (2015 March 11)&K2-3d (2015 Sept 12)\\
SMA/R$_*$&74.55$^{+6.81}_{-11.99}$&76.70$^{+11.99}_{-17.95}$\\
Incl (deg)&89.96$^{+0.42}_{-0.44}$&89.83$^{+0.79}_{-0.74}$\\
R$_p$/R$_*$&0.0240$^{+0.0008}_{-0.0008}$&0.0269$^{+0.0008}_{-0.0007}$\\
T0 (BJD)&2457093.57523$^{+0.00356}_{-0.00502}$&2457271.80073$^{+0.00194}_{-0.00169}$\\

Period (Spitzer Only)&\multicolumn{2}{c}{44.5563737$\pm$0.0011500\,d}\\\hline

Object&K2-26b (2015 March 12)&\\
SMA/R$_*$&23.578193$^{+1.580793}_{-3.422381}$&\\
Incl (deg)&89.223181$^{+1.063224}_{-1.039645}$&\\
R$_p$/R$_*$&0.050253$^{+0.001463}_{-0.001529}$&\\
T0 (BJD)&2457168.503765$^{+0.001777}_{-0.001801}$&\\

\enddata
\end{deluxetable}

\begin{deluxetable}{lcc|cc}
\tablecaption{Ephemeris Parameters From Combined \Ktwo and \Spitzer Transits\label{Periods}}
\tabletypesize{\scriptsize}
\tablehead{Planet &T0 (BJD)&Period (day)&$\Delta T0/\sigma^1$&$\Delta Period/\sigma^1$}
\startdata
K2-3b&2456813.42024$\pm$0.00094&10.054544$\pm$0.000029&	0.93&	2.0\\
K2-3c&2456812.2777$\pm$0.0026&24.64638$\pm$0.00018&-0.25&	0.74\\
K2-3d&2456826.2248$\pm$0.0038&44.55765$\pm$	0.00043&0.29&	-0.92\\
K2-26b&2456775.16503$\pm$0.00050& 14.568101$\pm$0.000020&-0.09&	0.88\\
\enddata
\tablecomments{$^1$Differences in T0 and Period from \Ktwo+\Spitzer to \Ktwo-only values from Table~\ref{K2data} relative to combined uncertainties. }
\end{deluxetable}

\begin{deluxetable}{lcc|cccc}
\tablecaption{Comparison of Planet/Star Ratio\label{Combined}}
\tabletypesize{\scriptsize}
\tablehead{Planet$^1$	&	Kepler	&	Spitzer	&	Diff/sigma	&	Kepler+Spitzer}	
\startdata
K2-3b Epoch1+2	&		&		&		&\\
Rp/R*	&	0.03483$\pm$0.00097	&	0.0353$\pm$0.0011	&	0.29&	0.0350$\pm$0.0007\\ 
Rp/R$_\oplus$ &		&		&		&2.08$\pm$0.25\\ \hline
K2-3c	&		&		&		&		\\
Rp/R*	&	0.027$\pm$0.001	&	0.02705$\pm$0.00005	&	0.04	&	0.0270$\pm$0.0006	\\ 
Rp/R$_\oplus$ &		&		&		&1.69$\pm$0.21\\ \hline
K2-3d	&		&		&		&		\\
Rp/R*	&	0.0250$\pm$0.0016	&	0.02545$\pm$0.0005	&	0.28	&	0.0252$\pm$0.0007	\\ 
Rp/R$_\oplus$ &		&		&		&1.61$\pm$0.20	\\ \hline
K2-26b	&		&		&		&		\\
Rp/R*	&	0.0471$\pm$0.0028	&	0.0412$\pm$0.0022	&	1.66	&	0.0434$\pm$0.0017	\\ 
Rp/R$_\oplus$ &		&		&		&2.47$\pm$0.40\\ \hline
\enddata
\tablecomments{$^1$Assumes R$_*$=0.561$\pm$ 0.068 R$_\odot$ for K2-3 \citep{Crossfield2015} and R$_*$=0.52$\pm$0.08 R$_\odot$ for K2-26. \citep{Schlieder2015}.}
\end{deluxetable}

\begin{deluxetable}{ccc|ccc|ccc}
\tablecaption{Central Transit Times for K2-3b\label{centertimes}}
\tabletypesize{\scriptsize}
\tablehead{Obs.	&	Orbit& BJD$^1$&Obs.	&	Orbit& BJD$^1$&Obs.	&	Orbit& BJD$^1$\\	
\multicolumn{3}{c}{K2-3b}&\multicolumn{3}{c}{K2-3b}&\multicolumn{3}{c}{K2-3b}}
\startdata
 \Ktwo &0&1980.4182$\pm$0.0015&\Ktwo&0&1979.2785$\pm$0.0026&\Ktwo&0&1993.2229$\pm$0.0039 \\
\Ktwo & 1 & 1990.4744$\pm$0.0016& \Ktwo &1 & 2003.9288$\pm$0.0028 &\Ktwo & 1 & 2037.7829$\pm$0.0049\\
 \Ktwo &2 & 2000.5259$\pm$0.0016 &\Ktwo &2 & 2028.5722$\pm$0.0037&\Spitzer & 6 & 2260.5752$\pm$0.0043 \\
 \Ktwo &3 & 2010.5817$\pm$0.0017&\Ktwo &3 & 2053.2151$\pm$0.0047& \Spitzer &10 & 2438.8007$\pm$ 0.0018 \\
 \Ktwo &4 & 2020.6334$\pm$0.0017 &\Spitzer &12 & 2275.0315$\pm$0.0028&& \\
 \Ktwo &5 & 2030.6876$\pm$0.0018 &\Spitzer & 19 & 2447.5598$\pm$0.0021&&\\
 \Ktwo &6 & 2040.7443$\pm$0.0020 &&&&\\
 \Ktwo &7 & 2050.7972$\pm$0.0017 &&&&\\
 \Spitzer &28 & 2261.94852$\pm$0.00070&&&&\\
 \Spitzer &29 & 2272.00231$\pm$0.00066 &&&&\\
 \Spitzer &46 & 2442.92813$\pm$0.00079 &&&&\\\hline
 \multicolumn{3}{c}{$\chi^2=23$ with 9 dof$^2$} &\multicolumn{3}{c}{$\chi^2=3.4$ with 4 dof}&\multicolumn{3}{c}{$\chi^2=1.4$ with 2 dof }\\
 \enddata
 \tablecomments{$^1$BJD-2454833. $^2$ The $\chi^2$ statistic for the hypothesis of a simple model with a constant period.}
\end{deluxetable}
\clearpage

\begin{thebibliography}




\bibitem[Almenara et al.(2015)]{Almenara2015} Almenara, J.~M., Astudillo-Defru, N., Bonfils, X., et al.\ 2015, \aap, 581, L7 

\bibitem[Batalha et al.(2014)]{Batalha2014} Batalha, N., Mandell, A., Kalirai, J., \& Clampin, M.\ 2014, Search for Life Beyond the Solar System.~Exoplanets, Biosignatures \& Instruments, P3 

\bibitem[Beichman et al.(2013)]{Beichman2013} Beichman, C., Ciardi, D., Akeson, R., et al.\ 2013, arXiv:1309.0918 

 \bibitem[Beichman et al.(2014)]{Beichman2014} Beichman, C., Benneke, B., Knutson, H., et al.\ 2014, \pasp, 126, 1134 

 
\bibitem[Berta-Thompson et al.(2015)]{Berta2015} Berta-Thompson, Z.~K., Irwin, J., Charbonneau, D., et al.\ 2015, \nat, 527, 204



\bibitem[Carter et al.(2012)]{Carter2012} Carter, J.~A., Agol, E., Chaplin, W.~J., et al.\ 2012, Science, 337, 556 


\bibitem[Claret \& Bloemen(2011)]{Claret2011} Claret, A., \& Bloemen, S.\ 2011, \aap, 529, A75 

\bibitem[Crossfield et al.(2015)]{Crossfield2015} Crossfield, I.~J.~M., Petigura, E., Schlieder, J.~E., et al.\ 2015, \apj, 804, 10 


\bibitem[Deck et al. (2014)]{Deck2014} Deck, K.~M., Agol, E., 
Holman, M.~J., \& Nesvorn{\'y}, D.\ 2014, \apj, 787, 132 

\bibitem[Deck \& Agol (2015)]{Deck2015} Deck, K.~M., \& Agol, E.\ 2015, \apj, 802, 116 



\bibitem[Deming et al.(2015)]{Deming15} Deming, D., Knutson, H., Kammer, J., et al.\ 2015, \apj, 805, 132

 \bibitem[Dressing \& Charbonneau(2013)]{Dressing2013} Dressing, C.~D., \& Charbonneau, D.\ 2013, \apj, 767, 95


 \bibitem[Dressing \& Charbonneau(2015)]{Dressing2015} Dressing, C.~D., \& Charbonneau, D.\ 2015, arXiv:1501.01623 

 \bibitem[D{\'e}sert et al.(2015)]{Desert2015} D{\'e}sert, J.-M., Charbonneau, D., Torres, G., et al.\ 2015, \apj, 804, 59 

\bibitem[Evans et al.(2015)]{Evans2015} Evans, T.~M., Aigrain, S., Gibson, N., et al.\ 2015, arXiv:1504.05942

 \bibitem[Fazio et al.(2004)]{Fazio2004} Fazio, G.~G., Hora, J.~L., Allen, L.~E., et al.\ 2004, \apjs, 154, 10 


\bibitem[Foreman-Mackey et al.(2013)]{Foreman2013} Foreman-Mackey, D., Hogg, D.~W., Lang, D., \& Goodman, J.\ 2013, \pasp, 125, 306 


\bibitem[Goodman \& Weare (2010)]{Goodman2010} Goodman, J. \& Weare, J. 2010, Comm. Appl. Math and Comp. Sci, 5, 65.

\bibitem[Greene et al.(2016)]{Greene2016} Greene, T.~P., Line, M.~R., Montero, C., et al.\ 2016, ApJ, 817, 17.

 \bibitem[Grillmair et al.(2012)]{Grillmair2012} Grillmair, C.~J., Carey, S.~J., Stauffer, J.~R., et al.\ 2012, \procspie, 8448, 84481I 



 \bibitem[Howard et al.(2012)]{Howard2012} Howard, A.~W., Marcy, G.~W., Bryson, S.~T., et al.\ 2012, \apjs, 201, 15 

 \bibitem[Howell et al.(2014)]{howell2014} Howell, S.~B., Sobeck, C., Haas, M., et al.\ 2014, \pasp, 126, 398 

\bibitem[Ingalls et al.(2012)]{Ingalls2012} Ingalls, J.~G., Krick, J.~E., Carey, S.~J., et al.\ 2012, \procspie, 8442, 84421Y 

\bibitem[Kreidberg et al.(2014a)]{KBD14a} Kreidberg, L., Bean, 
 J.~L., D{\'e}sert, J.-M., et al.\ 2014, \nat, 505, 69 (2014a)

\bibitem[Kreidberg et al.(2014b)]{Kreidberg2014b} Kreidberg, L., Bean, 
J.~L., D{\'e}sert, J.-M., et al.\ 2014, \apjl, 793, L27 

\bibitem[Krick et al.(2015)]{Krick2015} Krick, J., Ingalls, J. Carey, S. Grillmair, et al. 2015, {\it IRAC High Precision Photometry Website}, \url{http://irachpp.spitzer.caltech.edu}.

\bibitem[Knutson et al.(2014)]{Knutson2014} Knutson, H.~A., 
 Dragomir, D., Kreidberg, L., et al.\ 2014, \apj, 794, 155 
 
 
\bibitem[Knutson et al. (2012)]{Knutson2012} Knutson, H.A. Lewis, N., Fortney, J.J, 2012, \apj,754,22.


\bibitem[Line et al.(2013a)]{Line2013a} Line, M.~R., Wolf, A.~S., 
Zhang, X., et al.\ 2013, \apj, 775, 137 

\bibitem[Line \& Yung(2013b)]{Line2013b} Line, M.~R., \& Yung, Y.~L.\ 2013, \apj, 779, 3 



\bibitem[Mazeh et al.(2013)]{Mazeh2013} Mazeh, T.,
Nachmani, G.,  Holczer, T., et al.\ 2013, \apjs, 208, 16 

\bibitem[Mordasini et al.(2012)]{Mordasini2012} Mordasini, C., Alibert, Y., Benz, W., Klahr, H., \& Henning, T.\ 2012, \aap, 541, A97 




\bibitem[Parviainen(2015)]{Parviainen2015} Parviainen, H.\ 2015, \mnras, 450, 3233 

\bibitem[Payne \& Lodato(2007)]{Payne2007} Payne, M.~J., \& Lodato, G.\ 2007, \mnras, 381, 1597 


\bibitem[Ricker et al.(2014)]{Ricker2014} Ricker, G.~R., Winn, J.~N., Vanderspek, R., et al.\ 2014, \procspie, 9143, 914320 

\bibitem[Schlieder et al (2015)]{Schlieder2015} Schlieder, J.~E., Crossfield, I.~J.~M., Petigura, E.~A., et al. ,2016 \apj, 818,87.
 
\bibitem[Skrutskie et al.(2006)]{Skrutskie2006} Skrutskie, M.~F., Cutri, R.~M., Stiening, R., et al.\ 2006, \aj, 131, 1163 

\bibitem[Steffen et al.(2013)]{Steffen2013} Steffen, J.~H., Fabrycky, D.~C., Agol, E., et al.\ 2013, \mnras, 428, 1077 

\bibitem[Sullivan et al.(2015)]{Sullivan2015} Sullivan, P.~W., Winn, J.~N., Berta-Thompson, Z.~K., et al.\ 2015, arXiv:1506.03845, \apj, in press.
 

\bibitem[Weiss \& Marcy(2014)]{Weiss2014} Weiss, L.~M., \& Marcy, G.~W.\ 2014, \apjl, 783, L6 

\bibitem[Werner et al.(2004)]{Werner2004} Werner, M.~W., Roellig, T.~L., Low, F.~J., et al.\ 2004, \apjs, 154, 1 

\bibitem[Wong et al.(2015)]{Wong2015} Wong, I., Knutson, H.~A., Lewis, N.~K., et al.\ 2015, arXiv:1505.03158 

\bibitem[Wright et al.(2010)]{Wright2010} Wright, E.~L., 
 Eisenhardt, P.~R.~M., Mainzer, A.~K., et al.\ 2010, \aj, 140, 1868 


 \end{thebibliography}
\end{document}